\documentclass[12pt]{iopart}
\usepackage{graphicx}

\begin{document}

\title{Observation of an anomalous peak in isofield $M$($T$) curves in BaFe$_2$(As$_{0.68}$P$_{0.32}$)$_2$ suggesting a phase transition in the irreversible regime}
\author{ S. Salem-Sugui, Jr.$^1$, J. Mosqueira$^2$, A.D. Alvarenga$^3$, D. S\'o\~nora$^2$, E. P. Herculano$^1$, Ding Hu$^4$, Genfu Chen$^4$, Huiqian Luo$^4$ }
\address{$^1$Instituto de Fisica, Universidade Federal do Rio de Janeiro,
21941-972 Rio de Janeiro, RJ, Brazil}
\address{$^2$LBTS,Universidade de Santiago de Compostela, E-15782, Spain}
\address{$^3$Instituto Nacional de Metrologia Qualidade e Tecnologia, 25250-020 Duque de Caxias, RJ, Brazil}
\address{$^4$Beijing National Laboratory for Condensed Matter Physics, Institute of Physics, Chinese Academy of Sciences, Beijing  100190,  P. R. China}
\begin{abstract}
We measure magnetization as a function of temperature, magnetic field and time in a BaFe$_2$(As$_{0.68}$P$_{0.32}$)$_2$ single crystal with $T_c$ = 27.6 K. The fish-tail observed on many $M$($H$) curves are used to construct isofield $M$($T$) curves which show an anomalous peak at some temperature $T_t$ suggesting a possible phase transition in the irreversible regime. A vortex dynamics study performed along the peaks evidences a minimum in the relaxation rate occurring at the same position $T_t$ of the minimum value of $M$ in these peaks.  A vortex dynamics study performed on $M$($H$) curves show two distinct minimum in the relaxation rate: a first minimum ($H_1$) for a lower field correlating with $T_t$ and a second ($H_2$) correlating with the peak in the fish-tail. A phase diagram is constructed and the line corresponding to $T_t$ and $H_1$ points obey an expression developed in the literature for a structural  rhombic to square lattice phase transition, further supporting this view.
\end{abstract}
\maketitle

\section{Introduction}
Since the discovery of superconductivity in iron-pnictide systems \cite{japan,hu1,ren}, vortex dynamics has been the object of intensive study, as many different compounds belonging to this system show potential for applications mainly due to the considerably high-superconducting critical temperature, $T_c$,  high upper critical field $H_{c2}$, high critical current, and a considerably low anisotropy \cite{2,3,4}. Additionaly, iron-pnictides  also exhibit a considerable large flux-creep allowing to study in detail different pinning mechanisms. Among these studies, special attention has been given to the study of the second magnetisation peak (SMP), or fish-tail, which in pnictides shows similarities to those observed and studied in the high-$T_c$ superconductors.\cite{5,6,7,8} It is worth mentioning that the SMP was previously observed and studied in the low $T_c$ superconductors, as for instance in Nb.\cite{nb} The SMP is associated to a peak in the critical current highly interesting from the technological point of view for applications. The SMP, or fish-tail, appears on most single crystal pnictides as for instance in the 122 family BaFe$_2$(As,P)$_2$,\cite{kwok} Ba(Fe,Co)$_2$As$_2$,\cite{phaset,Bshen,naka} (Ba,K)Fe$_2$As$_2$,\cite{s1,Hyang} Ba(Fe,Ni)$_2$As$_2$,\cite{s2,s3} Ba(Fe,Ru)$_2$As$_2$,\cite{sharma}  and (Ba,Na)Fe$_2$As$_2$,\cite{prama2} in the 111 LiFeAs,\cite{prama} in the oxi-pnictides SmFeAs(O,F),\cite{hu,ding} CeFeAsO,\cite{shah} CeFeAs(O,F),\cite{chong} NdFeAsO$_{0.85}$,\cite{lesley} PrFeAsO$_{0.60}$F$_{0.12}$,\cite{DBhoi1,DBhoi} and in the chalcogenide Fe(Se,Te)\cite{marco,miu,Yadav} among others. As a result of the low anisotropy the SMP is observed for magnetic fields applied parallel, perpendicular or forming an angle with the c-axis of the sample \cite{s3}. The rich variety of explanations for the SMP, which for instance includes a vortex lattice phase transition \cite{phaset,prama}, a pinning crossover \cite{Bshen,s1}, an order disorder transition \cite{miu}, and a lack of evidence for pinning crossover or softness of the vortex lattice \cite{s2,s3}, may suggest that the mechanism responsible for the effect is sample dependent. \cite{s1,DBhoi} It is worth mentioning the important role that the multi-band character\cite{epl1,zehe}, along with the anisotropy and pinning, may play on the vortex dynamics of pnictides.\cite{prozorovSUST} Another important feature in pnictides is the existence of nano-scale variations in the gap as observed in gap-maps obtained through STM in Ba(Fe,Co)$_2$As$_2$ samples.\cite{hoffman1,hoffman2,massee} In that case, as a magnetic field is applied, one may expect that regions with a lower carrier density may become normal affecting the whole vortex distribution.

Among the explanations found for the SMP it is worth mentioning the phase transition of the vortex lattice which finger-print appears to be a minimum in the isofield relaxation rate -$R$ vs. $T$.\cite{phaset,prama,miu} This minimum  refers to a softness of the vortex lattice which due to energy considerations gives rise to a structural phase transition which leads to an increase in the relaxation rate as temperature increases, explaining the fish-tail form of the SMP.\cite{rosenstein1,rosenstein2,phaset} This well based explanation appear to be not in contradiction with the elastic to plastic pinning crossover observed near the peak field of the SMP, $H_p$.\cite{phaset}

In this work, we report on isothermic and isofield magnetisation data obtained on a single crystal of BaFe$_2$(As$_{0.68}$P$_{0.32}$)$_2$ with $T_c$ = 27.6 K and $\delta T_c$$\approx$ 1 K  for $H$$\parallel$ c-axis. Isothermial $M$($H$) curves obtained after zero-field cooling (zfc) show a pronounced second magnetisation peak which is apparent for temperatures from 3 K to 26 K. As a result of the pronounced fish-tail form of the curves, a plot of many $M$($H$) curves show several crossings occurring for many different magnetic fields, suggesting the existence of some anomaly on the isofield $M$($T$) curves in the irreversible regime. Isofield $M$($T$) curves are then obtained by extracting the magnetization values from $M$($H$) curves, and compared with  zfc $M$($T$) curves obtained by zero field cooling to a certain temperature $T$ at which the magnetic field is applied and data are collected as the temperature increases. While this last procedure gives rise to a smooth $M$($T$) curve, the one obtained from $M$($H$) measurements presents an anomalous peak, suggesting some sort of transition in the magnetisation.  Reproducibility of the anomalous peaks are checked by performing  additional magnetisation measurements $M$($H$,$T$) for temperatures running along the peaks where each data of these sets is obtained after a zfc procedure. To study the vortex dynamics along the anomalous $M$($T$) peaks, we measured the magnetisation as a function of time for each of these $M$($H$,$T$) measurements. The resulting isofield relaxation rates -$R$($T$) vs. $T$ plots show a minimun at approximately the same position at which the corresponding $M$($T$) data reaches a minimum suggesting that the anomalous peak is related to a vortex lattice phase transition. Magnetisation as a function of time is also obtained for fields running along the SMP on several $M$($H$) curves, allowing to study the different relaxation rate regimes. The resulting plots of $R$ vs. $H$ further corroborate the vortex lattice phase transition scenarious occurring below the SMP peak, $H_p$, while a pinning crossover is suggestive to occur above $H_p$. To our knowledge, the existence of an anomalous peak in the irreversible regime of $M$($T$) curves have not been observed before.

\section{Experimental}
The high-quality BaFe$_2$(As$_{0.68}$P$_{0.32}$)$_2$ single crystal used in this work (with approximated dimensions 4$\times$4$\times$0.3 mm$^3$ and mass $m$ = 3.582 mg) was grown by the BaAs/BaP flux method.\cite{Nakajima} The sample exhibits a sharp $T_c$ = 27.6 K with $\delta$$T_c$ $<$ 1K. Magnetic measurements were performed using a "MPMS-XL" system from Quantum Design (equipped with a magnetic shield) in both modes: the RSO mode was used for $M$($H$) measurements and DC scan were used for $M$($T$) measurements. The sample, as usual, was attached to a plastic straw allowing measurements with $H$$\parallel$c-axis. All measurements were obtained in a zfc procedure. Before lowering the temperature the magnetic shield is demagnetised and then the superconducting coil quenched, after which the remanent field is about 0.1 Oe. After cooling down from above $T_c$ to the desired temperature, the magnetic field is applied without overshooting and data is collected after field is declared stable. In the case of isothermal $M$($H$) hysteresis curves, data are collected with field increasing (or decreasing after a maximum field is reached) at fixed $\delta$$H$ increments. In that case we also obtained magnetic relaxation curves (over a period of 1-1.5 hours) for magnetic fields running along the SMP of the increasing field branch of selected $M$($H$) curves. In the case of $M$($T$,time) data for fixed fields, the magnetization as a function of time (over a period of 2 hours) is collected for each temperature after a zfc procedure. In the case of zfc isofield $M$($T$) curves data are collected with $T$ increasing at fixed $\delta$$T$ increments. Most of magnetic relaxation curves (not shown) present an initial transient stage with a comparatively low relaxation rate (observed before in Ba(Fe,Ni)$_2$As$_2$ system \cite{s2,s3})  holding for the first 10-15 minutes, after which the usual log($t$) behavior is achieved, allowing to extract the relaxation rate, defined as $R$ = dln($M$)/dln($t$). 

\section{Results and discussion}
\begin{figure}[t]
\includegraphics[width=\linewidth]{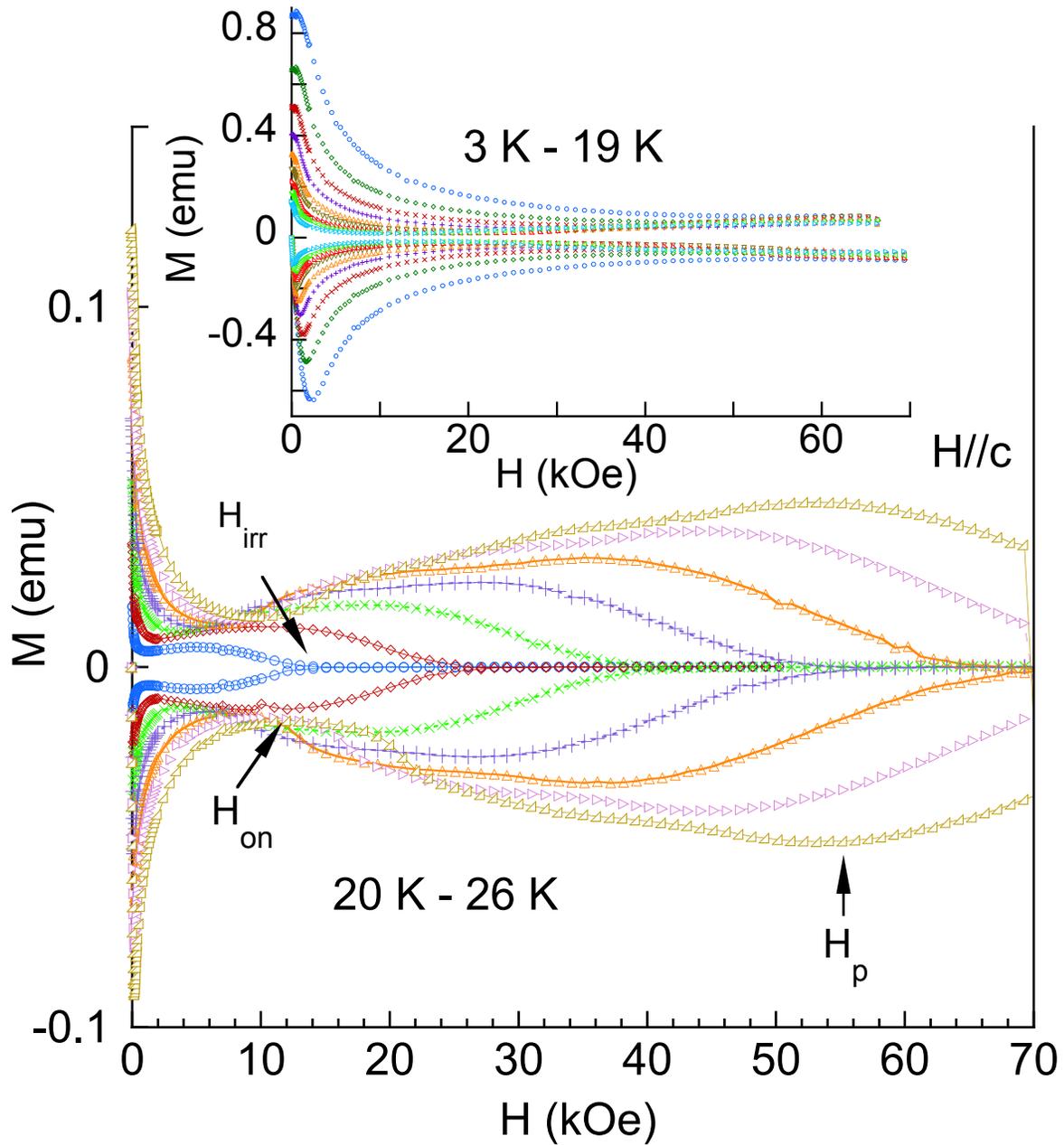}
\caption{Isothermal $M$($H$) curves as a function of temperature. Main: from 20-26 K in steps of 1K. Inset: from 3-19 K in steps of 2 K.}
\label{fig1}
\end{figure}
 \begin{figure}[t]
\includegraphics[width=\linewidth]{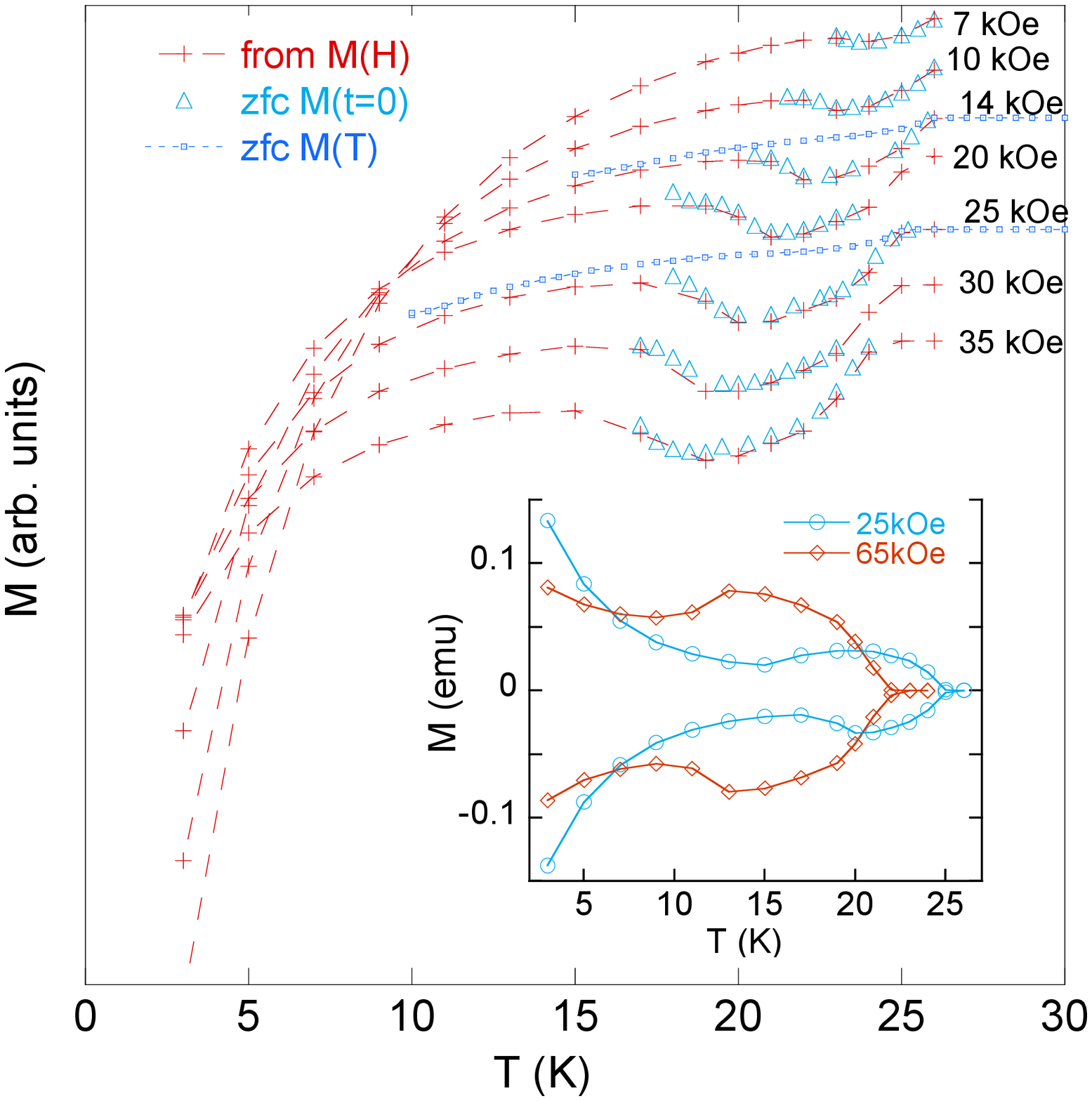}
\caption{Isofield $M$($T$) curves as obtained from 7-35 kOe in three different ways; red crosses - from $M$($H$) curves ; solid blue - corresponding to zfc $M$($T$) curves; and blue triangles corresponding to zfc $M$($t$=0) data, each one obtained in the same way as the $M$($H$) curves. The curves are dislocated along the y-axis for a better presentation. Inset: selected $M$($T$) curves obtained from both branches of the $M$($H$) curves.}
\label{fig2}
\end{figure}
Figure 1 shows isothermal $M$($H$) curves obtained for $H$$\parallel$ c-axis with temperatures running from 3 K to 26 K. The SMP is clearly visible for the curves above 20 K  from which is possible to identify the fields $H_{on}$, corresponding to the onset of the SMP, $H_p$, corresponding to the maximum of the SMP, and the irreversible field, $H_{irr}$. It is possible to see that  a much broader SMP takes place below 24 K and, as a consequence, it is difficult to precisely identify the position of $H_{on}$ below 20 K.  An interesting fact that can be visualized from the plots of Fig. 1 is the crossing of different $M$($H$) curves, which for a fixed $H$ leads to a maximum in $M$ upon increasing $T$.

To check this effect we obtained isofield $M$($T$) curves by extracting values of $M$ for fixed selected magnetic fields from the zfc $M$$(H)$ curves, which corresponds to a $M$($T$) curve with each point obtained after a zfc procedure. The resulting curves for selected magnetic fields obtained from the increasing field branch of the $M$($H$) curves are shown in Fig. 2, where the inset shows two selected curves with values of $M$ also extracted from the decreasing field branch of $M$($H$) curves, where the resulting curve interestingly resembles the fish-tail effect observed on $M$($H$) curves. It is interesting to see the existence of an anomalous peak in each curve of Fig. 2, with a minimum occurring at a temperature $T_t$. Since the anomalous peak develops for a fixed magnetic field (which may discard a pinning crossover) one may associate the peak to a possible phase transition occurring in the irreversible regime. To check for the reproducibility of the peaks we measured the zfc magnetization for temperatures running along each one of these peaks, where each data was taken after a zfc to the desired temperature, setting the magnetic field and measuring magnetization. To further study vortex dynamics, each one of these data was measured as a function of time. The resulting zfc $M$($t=0$) values are represented as open triangles in Fig. 2 and as shown, consistently reproduce the peaks in more detail. We also measured two isofield zfc $M$($T$) curves for $H$ = 14 and 25 kOe, which are plotted in Fig. 2. One can understand the absence of the peak in these zfc $M$($T$) curves as the $M$ values in this case do not correspond to the maximum allowed value in the irreversible regime. This fact may be a possible reason why this anomalous peak in $M$($T$) curves has apparently not been observed before, as the usual way to measure isofield zfc $M$($T$) curves is by continuously increasing the temperature. 
\begin{figure}[t]
\includegraphics[width=\linewidth]{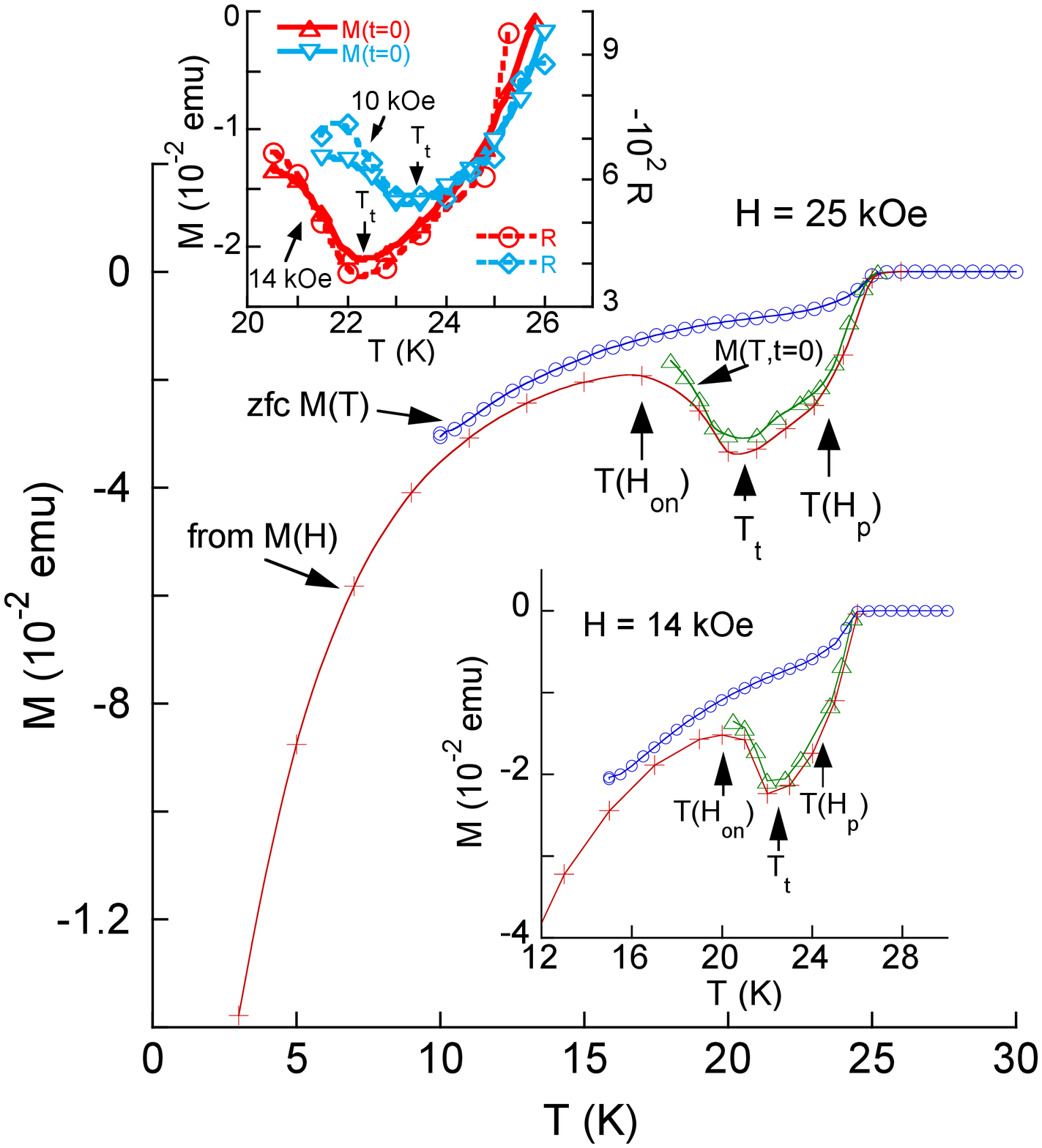}
\caption{Detail of the $M$($T$) curve for $H$ = 25 kOe. Lower inset:  detail of the $M$($T$) curve for $H$ = 14 kOe. Upper inset: double plots of $M$(t=0) vs. $T$ (open triangles in Fig. 3) and -$R$ vs. $T$ for $H$ = 10 and 14 kOe. The -$R$ vs. $T$ curves are shifted along the y-axis for a better presentation.}
\label{fig3}
\end{figure}

 Figure 3 shows the $M$($T$) curves obtained for $H$ = 25 kOe in the main figure and for 14 kOe in the lower inset where open triangles correspond to zfc $M$($t=0$) data obtained along the peaks for t=0 as discussed above. The plots in Fig. 3 allow to better visualize how the anomalous peak develops. Vortex dynamics along the anomalous peaks were studied by obtaining the relaxation rate $R$ of each $M$(time) data obtained along the peaks (open triangles). The upper inset of Fig. 3 shows -$R$ vs. $T$ plotted along with the corresponding $M$(t=0) vs. $T$ data for two selected magnetic fields, where it is possible to see that  -$R$ shows a minimum at a temperature at which virtually coincides with the position of the minimum in $M$(t=0) at $T_t$. This trend was observed for all -$R$ vs. $T$ curves. It is important to mention that the minimum in -$R$ occurring near $T_t$ in this inset may represent a softness of the vortex lattice which is followed by a steep increase of -$R$ as the temperature increases above $T_t$. According to Ref. \cite{phaset}, this steep increase of -$R$ explains the SMP which suggests that the anomalous peak is intrinsically related to the SMP observed in the $M$($H$) curves. To better exemplify the analogy between the anomalous peak and the SMP, we indicate in Fig. 3 with arrows the temperatures for which $H_{on}$ and $H_p$ are equal to the field in each plot. 
  \begin{figure}[t]
\includegraphics[width=\linewidth]{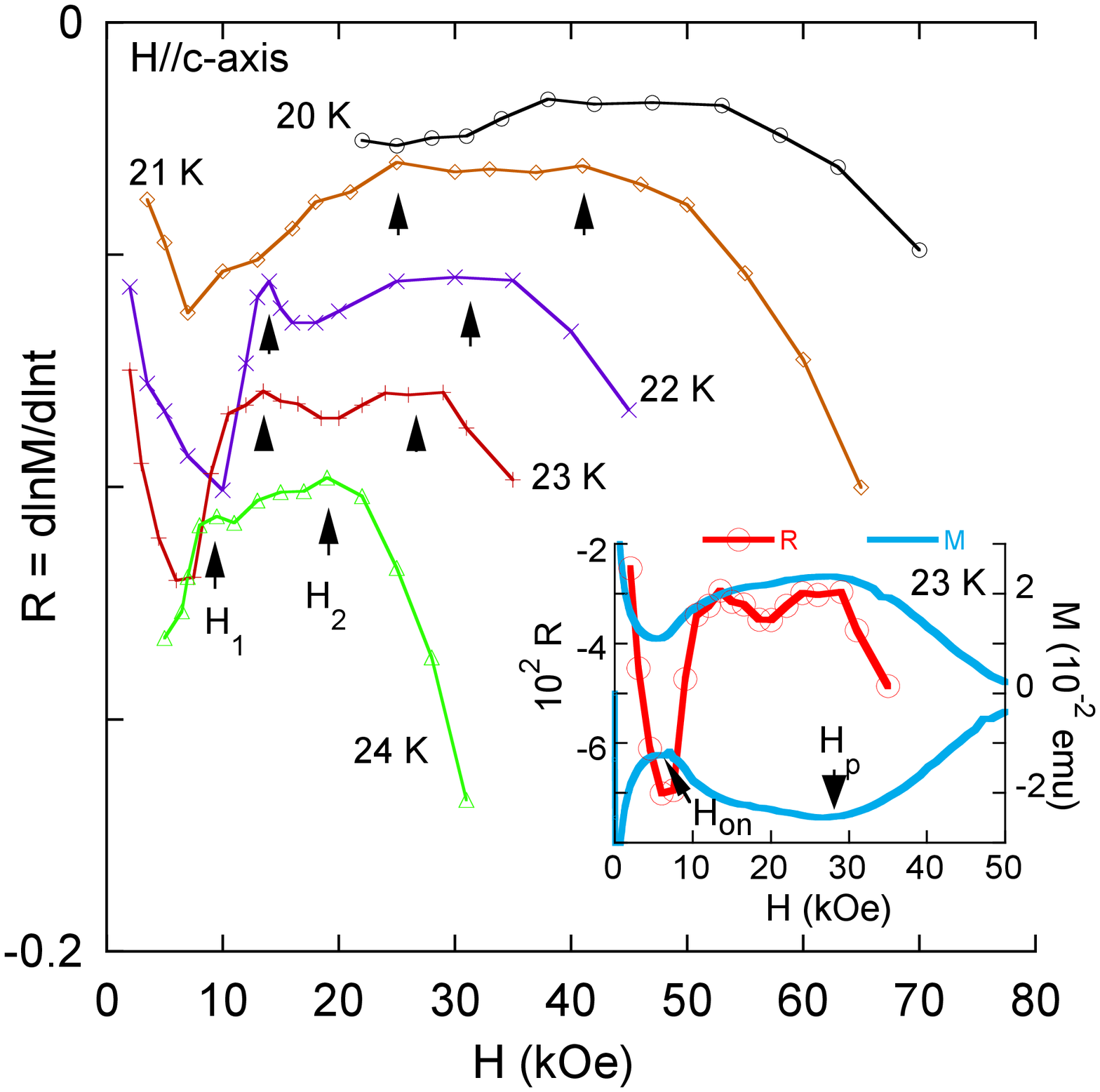}
\caption{Plots of $R$ vs. $H$ for $T$ = 20-24 K where the curves are dislocated along the y-axis for a better presentation (note that $|R|$ increases from top to botton). The inset shows a double plot of $R$ and $M$ vs. $H$ for $T$ = 23 K.}
\label{fig4}
\end{figure}
 \begin{figure}[t]
\includegraphics[width=\linewidth]{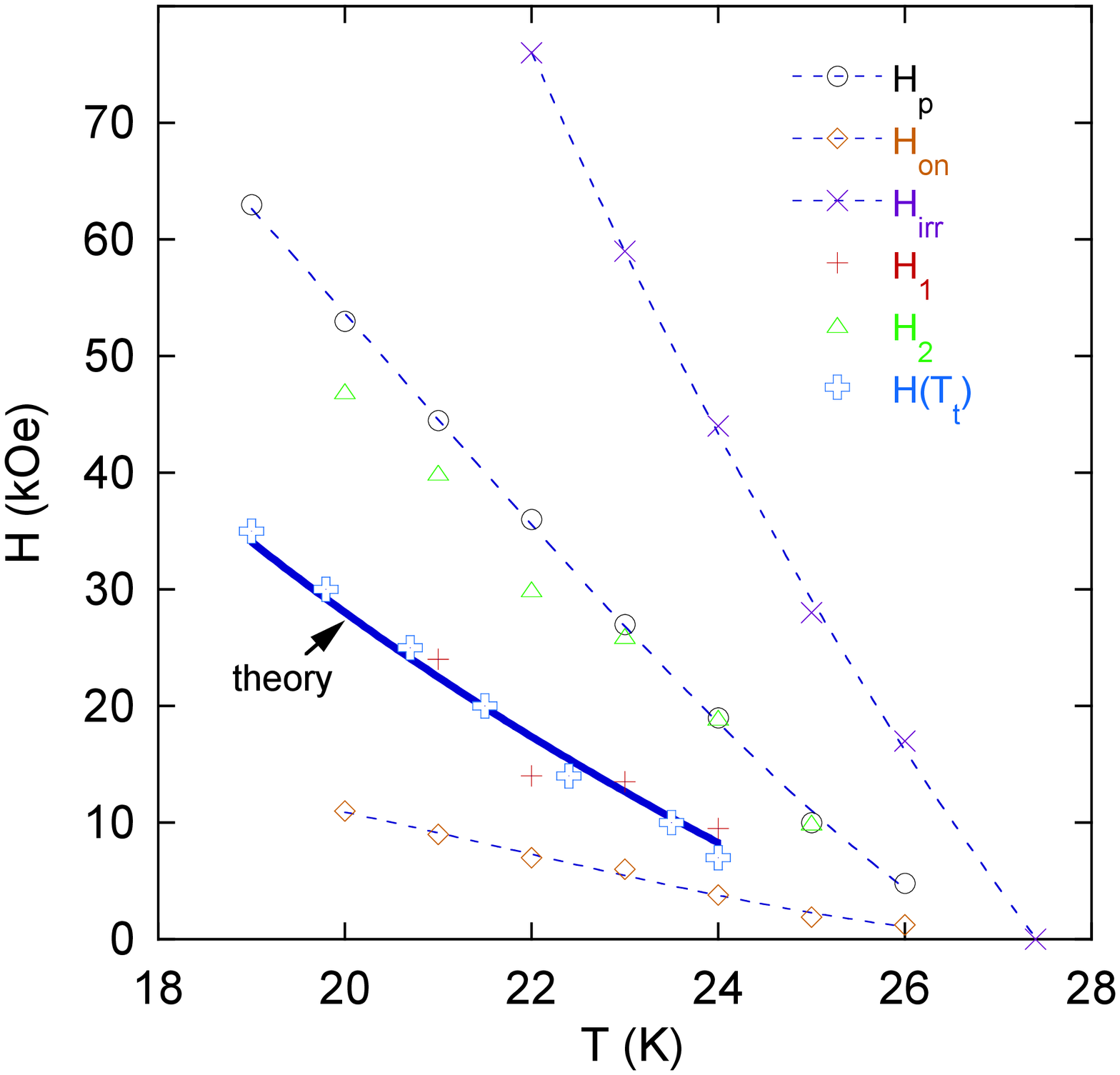}
\caption{Vortex phase diagram. The solid line is a fit to the theory (Eq.(1)]. Dotted lines are only guides to the eyes.}
\label{fig5}
\end{figure}

Figure 4 shows the results of $R$ vs. $H$ as obtained from magnetic relaxation data collected along the SMP on selected $M$($H$) curves. Interestingly, it is possible to identify a kind of double maximum (or minimum if one considers $|R|$) within the SMP for four (4) $M$($H$) curves. The maximum occurring between the fields $H_{on}$ and $H_p$, called $H_1$, appears to correlate with the peak in -$R$ vs. $T$, and the second maximum, called $H_2$, appears to be associated with $H_p$. This may suggest that the decrease in magnetization occurring above $H_p$ is associated with a change in the pinning mechanism, probably elastic to plastic, which as pointed in Ref.\cite{phaset} is not in contradiction with a vortex phase transition occurring below $H_p$. The inset of Fig. 4 shows a selected isothermal $R$ vs. $H$ curve plotted along with the corresponding $M$($H$) curve, where it is possible to visualise the positions of the double maximum with respect to the fields $H_{on}$ and $H_p$.

To sumarize the results, we plot the fields $H_{on}$, $H_p$, $H_{irr}$, $H_1$, $H_2$ and $T_t$ in a phase diagram in Fig. 5, where it is easy to visualise that $H_2$ is probably correlated with $H_p$, and $H_1$ with $T_t$, further suggesting that $T_t$ represents a phase transition of the vortex lattice in the irreversible regime.  It is worth mentioning that similar correlation between a minimum in -$R$ vs. $H$ with a minimum in -$R$ vs. $T$, where observed in Ref.\cite{phaset} for BaFe$_{2-x}$Co$_x$As$_2$ and the matching was associated to a vortex lattice phase transition as predicted and discussed in Refs.\cite{rosenstein1,rosenstein2}. This fact motivated us to fit the line formed by the points $T_t$ and $H_1$ in Fig. 5 to the expression presented in Ref.\cite{phaset} for a structural  rhombic to square lattice phase transition,
\begin{equation}
H_{spt}=A\frac{T_0-T}{C^{\nu -1}T^{\nu}},
\end{equation}
where $C$=$\frac{4\pi^3 \lambda^2}{L_z \phi_0^2}$, $\lambda$ is the London penetration depth, $L_z$ represents an effective superconducting layer width for thermal fluctuations \cite{phaset} and $\phi_0$ is the flux quantum.  We used the values $\lambda$ = 108 nm, which is appropriate for our sample\cite{kwok2} and $L_z$$\approx$ 10$^{-4}$ cm as in Ref.\cite{rosenstein1}. The fitting was conducted  by assuming $A$, $T_0$ and $\nu$ as free parameters. The resulting fitting, shown as a solid line in Fig.5, produces the values, $A$ = 0.98$T_c$, $T_0$ = 26.1 K and $\nu$ = 0.85. The values of the parameters $A$ and $T_0$ appear to be in reasonable agreement with the values found for Ba(Fe,Co)$_2$As$_2$\cite{phaset}  and La$_{2-x}$Sr$_x$CuO,\cite{rosenstein1} while a slightly smaller value was found here for the exponent $\nu$ which might be related to the smaller value of $\kappa$ = 47 for our sample\cite{kwok2} while $\kappa$ $\approx$ 75 for the samples in Refs.\cite{phaset,rosenstein1}. As mentioned above, there is an apparent absence of works in the literature showing a similar anomalous peak in $M$($T$) curves in the irreversible regime, which may indicate that the effect is only observed for the studied system. On the other hand, one month after this work was submitted, an interesting work was published in the arxiv.org data base\cite{kumar} where the authors observed a similar anomalous peak in the low $T_c$ superconductor Yb$_3$Rh$_4$Sn$_{13}$  evidencing that the effect could be of more general nature.
\section{Conclusions}
We observe an anomalous peak on isofield $M$($T$) curves in the irreversible regime of the studied sample which suggests a possible phase transition. To our knowledge, this anomalous peak has been observed for the first time in the present study. It is shown that the anomalous peaks with minimum $M$ value occuring at $T_t$ are associated to the SMP observed on $M$($H$) curves. Vortex dynamics studies performed along the peaks on $M$($T$) curves and along the SMP on $M$($H$) curves suggest that the anomalous peaks are related to a vortex lattice phase transition. The line formed by the points $H_1$ and $T_t$ (extracted from the anomalous peaks) is successfully fitted by a theory developed for  a structural rhombic to square lattice phase transition, further supporting this view.
\section*{Acknowledgements}
SSS, and ADA thanks support from the Brazilian agencies CNPq and FAPERJ. JM and DS acknowledge support from the Xunta de Galicia (grant no. GPC2014/038). The work at IOP, CAS, is supported by MOST (973 project: 2011CBA00110), NSFC (11374011and 91221303) and CAS (SPRP-B: XDB07020300).

\end{document}